\documentclass[pra,floatfix,twocolumn,reprint,longbibliography]{revtex4-1}
\usepackage{graphicx,bbm,amsmath,amssymb,units,xspace,subfigure}
\usepackage{siunitx}
\usepackage{xcolor}
\usepackage{hyperref}
\usepackage{braket}
\hypersetup{
    colorlinks=true,
    linkcolor=black,
    citecolor=black,
    filecolor=black,
    urlcolor=black,
}
\usepackage[capitalize]{cleveref}

\begin{document}
\title{Limits on the heralding efficiencies and spectral purities of spectrally-filtered single photons from photon pair sources}

\author{Evan Meyer-Scott}
\email{evan.meyer.scott@upb.de}
\affiliation{Integrated Quantum Optics, Department of Physics, University of Paderborn, Warburger Stra\ss e 100, 33098 Paderborn, Germany}
\author{Nicola Montaut}
\affiliation{Integrated Quantum Optics, Department of Physics, University of Paderborn, Warburger Stra\ss e 100, 33098 Paderborn, Germany}
\author{Johannes Tiedau}
\affiliation{Integrated Quantum Optics, Department of Physics, University of Paderborn, Warburger Stra\ss e 100, 33098 Paderborn, Germany}
\author{Linda Sansoni}
\affiliation{Integrated Quantum Optics, Department of Physics, University of Paderborn, Warburger Stra\ss e 100, 33098 Paderborn, Germany}
\author{Harald Herrmann}
\affiliation{Integrated Quantum Optics, Department of Physics, University of Paderborn, Warburger Stra\ss e 100, 33098 Paderborn, Germany}
\author{Tim J. Bartley}
\affiliation{Integrated Quantum Optics, Department of Physics, University of Paderborn, Warburger Stra\ss e 100, 33098 Paderborn, Germany}
\author{Christine Silberhorn}
\affiliation{Integrated Quantum Optics, Department of Physics, University of Paderborn, Warburger Stra\ss e 100, 33098 Paderborn, Germany}
\begin{abstract}
Photon pairs produced by parametric down-conversion or four-wave mixing can interfere with each other in multiport interferometers, or carry entanglement between distant nodes for use in entanglement swapping. This requires the photons be spectrally pure to ensure good interference, and have high heralding efficiency to know accurately the number of photons involved and to maintain high rates as the number of photons grows. Spectral filtering is often used to remove noise and define spectral properties. For heralded single photons high purity and heralding efficiency is possible by filtering the heralding arm, but when both photons in typical pair sources are filtered, we show that the heralding efficiency of one or both of the photons is strongly reduced even by ideal spectral filters with 100\% transmission in the passband: any improvement in reduced-state spectral purity from filtering comes at the cost of lowered heralding efficiency. We consider the fidelity to a pure, lossless single photon, symmetrize it to include both photons of the pair, and show this quantity is intrinsically limited for sources with spectral correlation. We then provide a framework for this effect for benchmarking common photon pair sources, and present an experiment where we vary the photon filter bandwidths and measure the increase in purity and corresponding reduction in heralding efficiency. 
\end{abstract}

\maketitle
\section{Introduction}
Photon pairs from nonlinear optics are so far the only resource to have distributed quantum entanglement over more than a few kilometers~\cite{ursin-2007-3,1367-2630-11-8-085002,Dynes:09,Inagaki:13,Ma2012Quantum-,Herbst17112015}, a critical link in future quantum networks, and are well-suited for use in multi-port quantum interferometers for sensing, simulation and computation, both as pairs directly and for heralded single photons~\cite{Metcalf:2013aa,Crespi:2013aa,Tillmann:2013aa,Carolan14082015}. Entangled photon pairs have also been used in quantum teleportation~\cite{Takesue:15,Valivarthi:2016aa,Sun:2016aa} and entanglement swapping~\cite{Takesue:09,PhysRevA.85.032337,Jin:2015aa}. These applications require that the reduced spectral state of each photon is pure: mixedness of the photon states leads to reduced visibility of the interference of independent photons, and therefore lower-quality final states. 

Parametric down-conversion (PDC) and four-wave mixing (FWM) are the most common sources of photon pairs, and these photons usually possess spectral anti-correlation leading to mixedness of the reduced state of each photon. This frequency entanglement can be useful for some applications~\cite{PhysRevA.65.053817}, but is catastrophic for multi-photon interference or entanglement-swapping experiments. A convenient solution is narrowband filtering of both photons, which casts each into a single spectral mode, removing entanglement in favor of the spectral purity of each photon. Both FWM sources~\cite{Sharping:06,1367-2630-13-6-065005,chalco2011,W.2013On-chip-,Sun:2016aa} and PDC sources ~\cite{PhysRevA.81.021801,Bruno:2014aa,Valivarthi:2016aa,PhysRevLett.117.210502,Chen:17,Vergyris:2016aa} often use filters much narrower than the photon bandwidths. But is spectral filtering compatible also with high {\em pair-symmetric heralding efficiency} (PSHE), defined as the product of signal and idler heralding efficiencies? In contrast to heralded single photon sources where only one photon requires high heralding efficiency, we consider photon pair sources where both photons must be generated in spectrally pure states and with high efficiency, such that both may be used for interference experiments. High heralding efficiency is critical for scaling experiments and communications to many photons and higher rates~\cite{1367-2630-12-9-093027,PhysRevLett.117.210502,Chen:17} due to the exponential increase in losses with number of photons, and also of fundamental importance: for reaching scalability in optical quantum computing~\cite{PhysRevLett.100.060502,PhysRevA.81.052303,Jennewein:2011aa}, in device-independent quantum cryptography~\cite{PhysRevLett.98.230501,PhysRevA.86.032325}, and for tests of local causality with entangled photons~\cite{LHFVienna,PhysRevLett.115.250402}. Our results are especially important for applications that require both high pair-symmetric heralding efficiency and multi-source interference: interference of pair sources to produce large entangled states~\cite{Pan2000Experime,1367-2630-12-9-093027,PhysRevLett.117.210502}, entanglement swapping~\cite{PhysRevLett.80.3891,Takesue:09,PhysRevA.85.032337,Herbst17112015,Jin:2015aa}, heralded noiseless qubit amplification~\cite{Kocsis2013Heralded,2015arXiv150703210B}, quantum repeater networks~\cite{Azuma:2015aa,PhysRevA.92.022357,Krovi:2016aa}, and certain multiphoton phase estimation schemes~\cite{0295-5075-82-2-24001,Xiang:2013aa}.

Here we show that, for photon pair sources with spectral correlation or anti-correlation, increasing the spectral purity by filtering comes at a direct cost of decreasing the pair-symmetric heralding efficiency. This tradeoff is based only on the joint spectral intensity (JSI) of the photons, not on the underlying physics that produce a specific JSI, meaning our results are applicable to both PDC and FWM, and to pulsed and continuous-wave pumps. We find a significant drop in achievable PSHE even with ideal filters. We quantify this tradeoff by introducing the symmetrized fidelity of the photon pairs to two spectrally pure single photons, and show that it is bounded well below one for spectrally-correlated sources. This is supported by an experiment using a lithium niobate photon-pair source, where we vary filter parameters, and find that heralding efficiency necessarily decreases as purity increases. Similar results could be obtained for spatial correlation and spatial filtering, but here we focus on a single spatial mode.

Previous investigations of filtering in PDC and FWM have largely focused on heralded single photons, where the herald{\em ing} photon is strongly filtered and the herald{\em ed} photon is unfiltered, allowing both high spectral purity and high single-sided heralding efficiency~\cite{Aichele:2002aa,Neergaard-Nielsen:07,1367-2630-12-6-063001,PhysRevA.83.053843}. The effect filtering on continuous-variable photon states has been studied~\cite{PhysRevA.90.023823}, as has the effect of self-and cross-phase modulation on filtered photon pairs~\cite{PhysRevA.94.063855}. Recent theoretical work has included also spatial entanglement and purity with spatial and spectral filters~\cite{PhysRevA.91.013819,PhysRevA.94.069901}, showing again high single-sided heralding efficiency and purity. This is in contrast to source engineering methods, which achieve intrinsically pure states by controlling the dispersion and pump bandwidth~\cite{PhysRevA.56.1627,PhysRevA.64.063815,PhysRevLett.100.133601,prespw,Garay-Palmett:07,Halder:09,Levine:10,PhysRevLett.106.013603,Gerrits:11,Fang:13,Harder:13,Fortsch:2013aa,Bruno:14,Weston:2016aa}. Some schemes with tight spectral and time filtering can even outperform this source engineering, when considering production rates as well as purity~\cite{PhysRevA.82.043826,PhysRevA.84.033844}. Furthermore, in contrast to spectral filtering after generation, placing the nonlinear medium in a cavity of carefully engineered length and finesse can in principle produce spectrally pure states without loss of heralding efficiency~\cite{Chuu_2012,1367-2630-17-7-073039}. In most cases, however, filters are still needed for single-mode operation, as the phasematching bandwidth covers multiple longitudinal modes of the cavity~\cite{Jeronimo-Moreno2010,Neergaard-Nielsen:07,PhysRevLett.101.190501,PhysRevLett.102.063603,PhysRevLett.110.220502}; fortunately, for narrowband pumps and filters, these modes do not contribute to a decrease in heralding efficiency because each filter intersects just one cavity mode.  For the case where both photons are to be used from non-engineered and non-cavity sources, hints that filtering is incompatible with high PSHE have appeared numerous times~\cite{PhysRevA.56.1627,PhysRevA.64.063815,PhysRevLett.100.133601,Bock:16} and a simple model for heralding efficiency after filtering was developed in~\cite{PhysRevA.92.012329}, but so far no experiments have directly studied the impact of filtering on purity and heralding efficiency simultaneously, and no previous studies have found the fundamental limits to symmetrized fidelity we present here.

\section{Spectrally-filtered photon pairs}

\begin{figure}[!h]
\centerline{\includegraphics[width=0.9\linewidth]{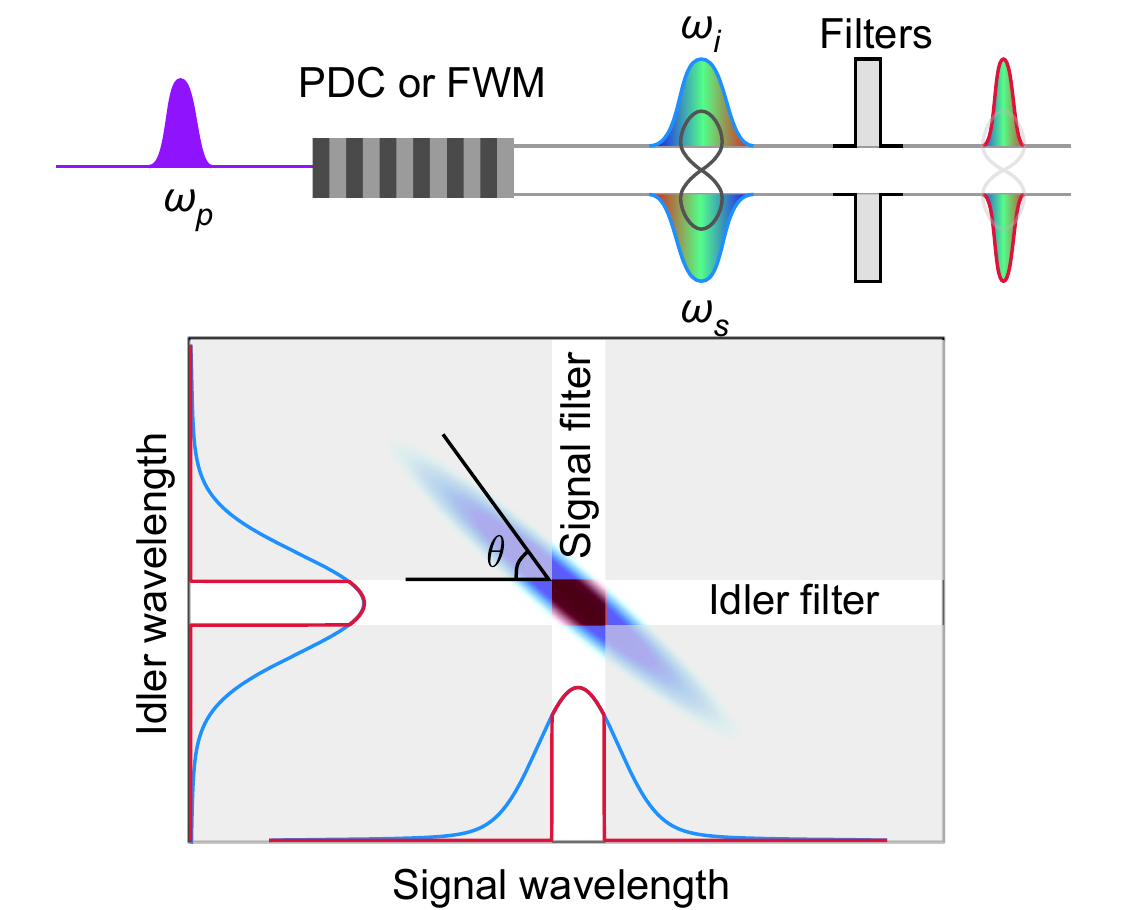}}
\caption{Photon pair production and filtering (top), resulting in joint spectral intensity with spectral correlation between signal and idler photons (bottom). The signal and idler filters are overlaid, and the JSI and marginal photon spectra remaining after filtering dictate the reduced-state spectral purity and heralding efficiency of the photons. The phasematching function with angle $\theta$ is multiplied by the pump envelope (which always has angle \SI{45}{\degree}) to produce the total JSI. Thus the overall angle of the JSI is somewhere between \SI{45}{\degree} and $\theta$.}
\label{fig.intro}
\end{figure}

One can get a feeling for the intrinsic tradeoff between reduced-state spectral purity and heralding efficiency from \cref{fig.intro}. It shows the joint spectral intensity of an example photon pair state, overlaid with narrowband filters on each photon, labeled signal and idler. To achieve a spectrally pure state, the JSI that remains after filtering must be uncorrelated between the two photons, either a circle, or an ellipse along the vertical or horizontal axis. But for high PSHE, the two-photon amplitudes transmitted by each filter individually must overlap, otherwise signal photons will pass the filter without the corresponding idler and vice versa.

An uncorrelated JSI, fully contained within both filters is only possible for certain ranges of the phasematching angle, namely $\theta\in\left[\SI{90}{\degree},\SI{180}{\degree}\right]$, and with a pump bandwidth optimized for the phasematching bandwidth. But these conditions are precisely those for which filtering is not required, since there are no underlying spectral correlations in this condition. Furthermore, achieving a phasematching angle in this range is nontrivial, as it requires the group velocity of the pump to be between that of the signal and idler. This source engineering is only possible~\cite{Zhang2012Heralded} in PDC for very specific wavelength ranges in birefringent crystals~\cite{prespw,1367-2630-10-9-093011,Laudenbach:16}. It is easier to arrange in FWM since it occurs naturally for normal dispersion with the pump between the signal and idler frequencies (here the frequencies are rather close, necessitating narrow filtering for pump removal), or by pumping near the zero-dispersion wavelength~\cite{PhysRevLett.102.123603,1367-2630-13-6-065009} or using birefringent fibers~\cite{Smith:09}.

As a concrete example we consider waveguided type-II PDC wherein the photons are emitted in a single spatial mode (such that spatial variables do not play a role) but with different polarizations. These sources can be easily transformed to entangled-pair sources with Sagnac~\cite{PhysRevA.73.012316} or Mach-Zehnder~\cite{PhysRevLett.105.253601} interferometers. At low enough pump powers to stay in the single-pair regime, the spectral properties of PDC are governed by the joint spectral amplitude $f\left(\omega_s, \omega_i\right)$ for signal and idler frequencies $\omega_s$ and $\omega_i$, giving rise to the photon pair state~\cite{PhysRevA.56.1627} 
\begin{align}
\ket{\psi} = \iint d\omega_s d\omega_i f\left(\omega_s, \omega_i\right) \mathcal{F}_s(\omega_s)\mathcal{F}_i(\omega_i)\ket{\omega_s}\ket{\omega_i},
\label{eq.state}
\end{align}
where $\ket{\omega_{s/i}}$ is a single photon at frequency $\omega_{s/i}$ with the polarization of the signal/idler mode and $\mathcal{F}_s(\omega_s)$ and $\mathcal{F}_i(\omega_i)$ are spectral filters on the signal and idler photons respectively. The joint spectral intensity is $\left|f\left(\omega_s, \omega_i\right)\mathcal{F}_s(\omega_s)\mathcal{F}_i(\omega_i)\right|^2$, and the filters can be of any shape: we consider square and Gaussian filters.

We model the joint spectral amplitude around central frequencies $\omega_{s0}$ and $\omega_{i0}$ by
\begin{align}\label{eq.jsa}
f\left(\omega_s, \omega_i\right) = N \exp\left({\frac{-\left(\omega_s-\omega_{s0}+\omega_i-\omega_{i0}\right)^2}{4\sigma_p^2}}\right)\\
\times\mathrm{sinc}\left(\frac{\left(\left[\omega_s-\omega_{s0}\right]\sin{\theta} + \left[\omega_i-\omega_{i0}\right]\cos{\theta}\right)}{2\sigma_{pm}}\right)\nonumber.
\end{align}
The pump and phasematching bandwidths are $\sigma_p$ and $\sigma_{pm}$, respectively; $N$ is a normalization term; and the phasematching angle~\cite{Smith:09} is $\theta = \arctan\left(\frac{k_p^\prime - k_s^\prime}{k_p^\prime - k_i^\prime}\right)$,
where $k_x^\prime$ is the frequency derivative of the wavenumber $k$ of mode $x$. Thus the nonlinear material, waveguide characteristics, and wavelengths can all be chosen to determine the phasematching angle.

\section{Heralding efficiency and reduced-state spectral purity}
We define the signal photon's {\em filter heralding efficiency} as the probability that the signal photon passes its filter given that the idler photon has passed its filter, and vice versa for the idler photon's filter heralding efficiency. These efficiencies will be less than one whenever the JSIs passed by each filter individually do not match~\cite{PhysRevA.91.013819,PhysRevA.92.012329}. Defining the probability that both photons pass their filters as $\Gamma_{both}$ and the probability that each passes individually as $\Gamma_{s/i}$, we find the signal's filter heralding efficiency is $\eta_{f,s} = \frac{\Gamma_{both}}{\Gamma_{i}}$,
and the idler's is $\eta_{f,i} = \frac{\Gamma_{both}}{\Gamma_{s}}$. Then we define the pair-symmetric heralding efficiency as $PSHE = \eta_{f,s}\eta_{f,i}$. Of course this is only the contribution of filtering to the PSHE; optical losses will lower the PSHE further.

The spectral purity of the reduced state of either photon given that both photons have passed their respective filters (corresponding to the relevant case of coincident detection) is~\cite{1367-2630-12-11-113052} $P = \mathrm{Tr}\left(\rho_s^2\right)$, where 
\begin{align}\label{eq.rhos}
\rho_s&=\mathrm{Tr}_i\left(\ket{\psi}\bra{\psi}\right)\\\nonumber
& = \iiint d\omega_i d\omega_s d\omega_s^\prime f\left(\omega_s, \omega_i\right)f^*\left(\omega_s^\prime, \omega_i\right)\\\nonumber
&\times \mathcal{F}_s(\omega_s)\mathcal{F}_s(\omega_s^\prime)\mathcal{F}_i(\omega_i)^2\ket{\omega_s}\bra{\omega_s^\prime}
\end{align} is the reduced density matrix. The purity can be taken for either signal or idler as there is no other degree of freedom (e.g. spatial) that would allow different purities for each mode and we are always considering that the photons are detected in coincidence. 
\begin{figure}[h]
\centerline{\includegraphics[width=\linewidth]{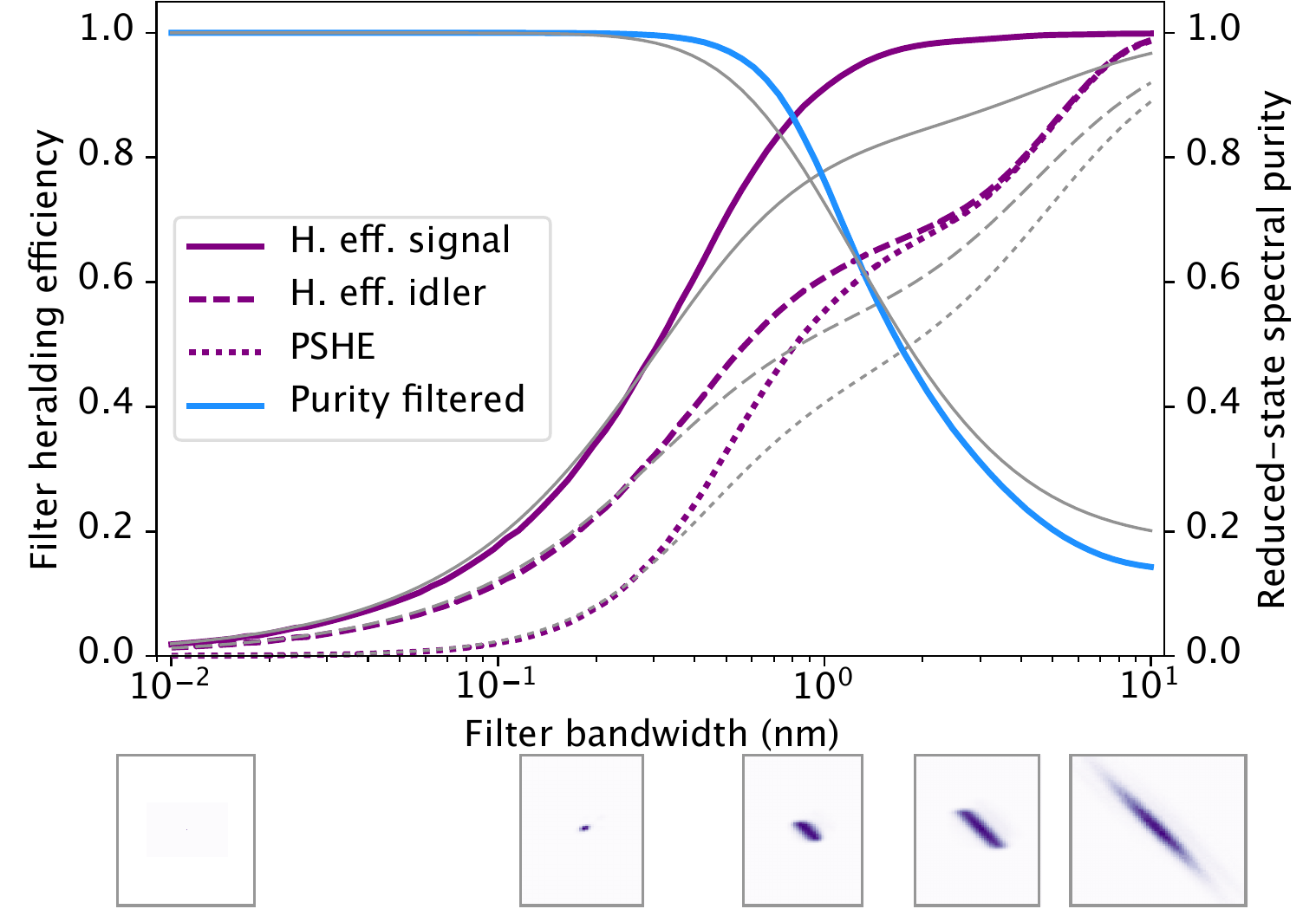}}
\caption{Theoretical filter heralding efficiency for signal (purple solid) and idler (purple dashed), combined PSHE (purple dotted), and spectral purity (blue) versus filter bandwidth, for the flat-top filters with the same bandwidth for signal and idler, showing the intrinsic tradeoff between purity and efficiency. The corresponding thin grey curves are the analytic results for Gaussian filters. Some representative JSIs are shown below their corresponding filter bandwidths (the leftmost is very small on this scale).}
\label{fig.filt}
\end{figure}

Taking the JSI of \cref{fig.intro} (with pump bandwidth \SI{0.42}{\nano\meter}, phasematching bandwidth \SI{0.46}{\nano\meter}, and $\theta=\SI{60.5}{\degree}$, matching the experiment below), we calculate the filter heralding efficiencies and spectral purity versus filter bandwidth, which are taken as equal for the signal and idler. As seen in \cref{fig.filt}, as soon as the filters are narrow enough to increase the purity, the filter heralding efficiency starts to drop. The filters are ideal flat-top filters with perfect transmission in the passband and perfect blocking otherwise. This is an idealization of real dense-wave-division multiplexing filters, chosen to highlight the intrinsic physical effects of filtering rather than the technical effects. In fact, real filters lead to even stronger reductions in heralding efficiency due to nonuniformities, slow rolloff, and nonunit transmission. Gaussian filters (thin grey curves) show worse performance for both purity and heralding efficiency, with the improved purity at large filter bandwidths due to the removal of sinc lobes under the Gaussian approximation of the JSI. The kink in \cref{fig.filt} around \SI{3}{\nano\meter} filter bandwidth in the idler heralding efficiency is due to the asymmetry of the JSI~\cite{0953-4075-46-5-055501}. Even though both filters are varied equally, since the JSI is tipped slightly towards parallel to the idler axis, above the kink, the filtering is dominated by the idler filter, while below both filters contribute.

\begin{figure}[h]
\centerline{\includegraphics[width=\linewidth]{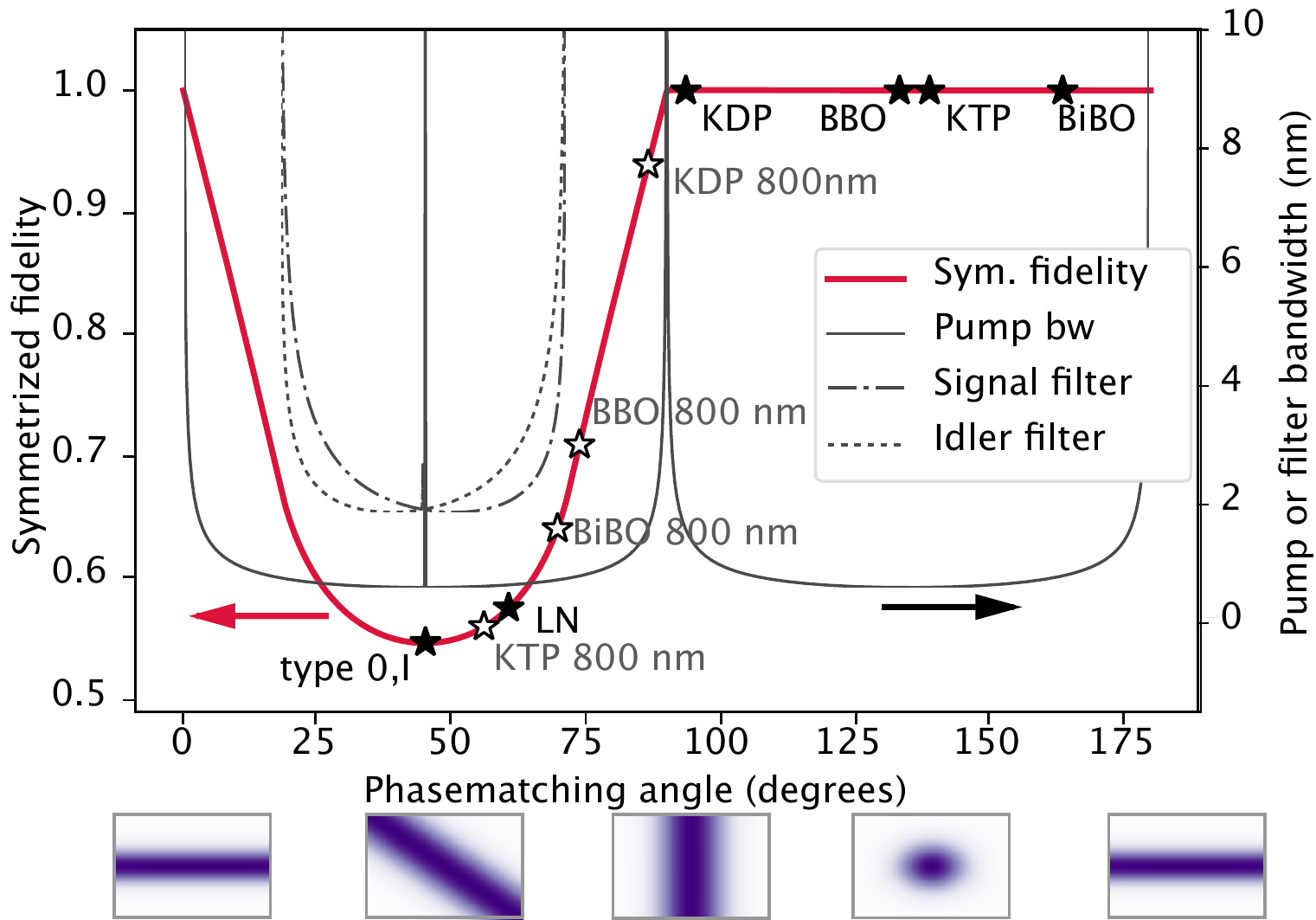}}
\caption{Calculated symmetrized fidelity $\sqrt{F_sF_i}$ (thick red) versus phasematching angle, after optimizing the pump bandwidth (black solid) and the signal (black dot-dash) and idler (black dash) filter bandwidths. The maximum achievable fidelity is independent of the crystal length but the optimal values of bandwidth change to accommodate the different phasematching bandwidths. A few crystal types~\cite{snlo} for degenerate type II PDC to \SI{1550}{\nano\meter} are overlaid (filled star), while for degenerate type 0 and type I, the phasematching angle is always \SI{45}{\degree} (except with engineered dispersion for example in microsctructured fibers~\cite{Garay-Palmett:07}, or for noncollinear PDC~\cite{1367-2630-12-9-093027}). With nondegenerate photons and other wavelengths (see three examples at \SI{800}{\nano\meter}, empty star) many different angles can be reached~\cite{Laudenbach:16}. Below the plot are unfiltered JSIs at \SI{45}{\degree} intervals.}
\label{fig.theta}
\end{figure}

To quantify the combined effect of filtering on heralding efficiency and purity we introduce the symmetrized fidelity $F=\sqrt{F_sF_i}$, where $F_{s/i}$ is the fidelity for the signal/idler to a pure single-photon state $\ket{1}_{s/i} = \int d\omega g_{s/i}(\omega)\hat{a}_{s/i}^\dagger(\omega) \ket{0}$ after heralding by the idler/signal and including the vacuum component caused by filtering losses. We symmetrize the fidelity in this way rather than taking just the signal or idler fidelity to capture the effects of filtering on both photons together. The spectral function $g_{s/i}(\omega)$ is optimized for each photon to maximize the fidelity, as it is not directly given by any eigenvector of the reduced density matrix~\cref{eq.rhos}. The individual fidelities are 
\begin{align}
F_{s} &=\eta_{f,s}\times\underset{g_s(\omega)}{\max}\bra{1}_s\rho_{s}\ket{1}_{s},\\\nonumber
F_{i} &= \eta_{f,i}\times\underset{g_i(\omega)}{\max}\bra{1}_i\rho_{i}\ket{1}_{i}.
\end{align}
Either $F_s$ or $F_i$ can  be made to approach one by filtering, but in general not both simultaneously. Using the Gaussian approximation developed in the Supplemental Material which allows analytic solutions, we find the symmetrized fidelity to be related to the purity and heralding efficiency by
\begin{align}
F = \sqrt{\eta_{f,s}~\eta_{f,i}}~\frac{2P}{1+P}.
\end{align}

By optimizing the pump and filter bandwidths for each phasematching angle we bound the maximum value of symmetrized fidelity available by filtering, as shown in \cref{fig.theta}. The maximum is independent of the phasematching bandwidth (here chosen as \SI{1.5}{\nano\meter}), though the optimal pump and filter bandwidths change. For our lithium niobate (LN) crystal with $\theta=\SI{60.5}{\degree}$ the maximum is $F=0.57$. By contrast, sources with $\theta\in\left[\SI{90}{\degree},\SI{180}{\degree}\right]$ can have $F\rightarrow 1$ even without filtering, as the optimal filter bandwidth goes to infinity. This shows clearly the futility of filtering for reduced-state spectral purity in PDC: the conditions in which filters are needed are only where filtering cannot recover perfect fidelity due to lowered heralding efficiency. Of course without filters in these conditions the fidelity to a pure single photon would be even lower. We stress that this fidelity bound is generic for all PDC and FWM sources (with JSIs described by the pump-times-phasematching model), and is thus a very powerful  tool in source design.

\begin{figure}[htp]
\centerline{\includegraphics[width=\linewidth]{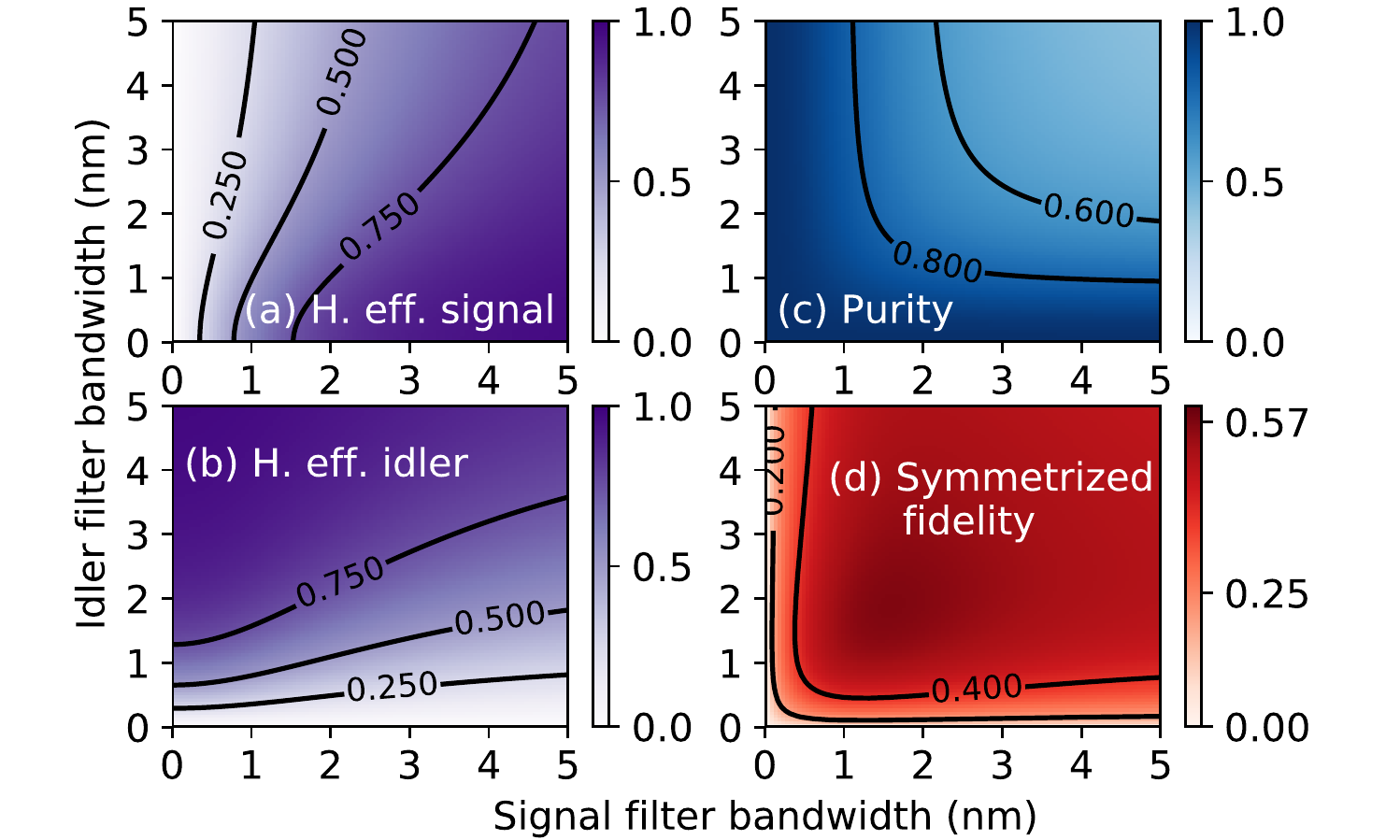}}
\caption{Filter heralding efficiencies for (a) signal and (b) idler as a function of signal and idler filter bandwidths, along with (c) reduced-state spectral purity and (d) symmetrized fidelity to a spectrally pure single photon. Here we use Gaussian filters, with bandwidths given by the full width at half-maximum (FWHM). While the heralding efficiencies and purity range individually over $[0,1]$, the symmetrized fidelity is reasonably constant around its mean value of 0.45, and never surpasses 0.57.}
\label{fig.filters}
\end{figure}

Finally, to show the sharpness of these effects we vary the filter bandwidths independently and set the pump and phasematching bandwidths to \SI{0.38}{\nano\meter} and \SI{1.5}{\nano\meter} respectively, which for $\theta=\SI{60.5}{\degree}$ allows an optimal symmetrized fidelity. As shown in \cref{fig.filters}, the best filter heralding efficiencies for the signal photon have the largest signal filter and the smallest idler filter; and vice versa for the idler photon. However the largest purity requires small filters on both arms, resulting in a symmetrized fidelity that varies slowly over filter bandwidth and never exceeds 0.57, falling to zero as either filter gets too narrow.

\section{Experiment}\label{sec.exp}

To confirm the tradeoff between purity and PSHE, we measured the heralding efficiency of signal and idler photons and the joint spectral intensities of a photon-pair source under various filtering conditions. The source (\cref{fig.intro}) was a \SI{21}{\milli\meter} type-II periodically-poled lithium niobate waveguide, fiber pigtailed on both ends~\cite{al.:2016aa} and pumped by a  Ti:Sapphire pulsed laser of wavelength \SI{778}{\nano\meter}. The laser had a pulse width of \SI{3.0}{\pico\second} FWHM, nearly transform limited to \SI{0.42}{\nano\meter} FWHM spectral bandwidth, and \SI{5}{\micro\watt} coupled power resulting in a production of  $\sim\num{0.02}$~pairs/pulse before filtering. Calculations for lithium niobate predict a phasematching angle of \SI{60.5}{\degree} and bandwidth \SI{0.46}{\nano\meter}. The output of the source was coupled to a WaveShaper 4000 (Finisar Corp.) which was used to separate the nondegenerate photons (central wavelengths \SI{1562}{\nano\meter} and \SI{1549}{\nano\meter}) and define their spectral filters.

We characterize the heralding efficiency for each filter setting using the Klyshko method~\cite{0049-1748-10-9-A09} such that 
$\eta_s = \frac{C}{S_i},~~\eta_i = \frac{C}{S_s}$,
where $C$ are the number of coincidences, $S_{s/i}$ are the number of signal/idler singles, and $\eta_{s/i}$ are the total heralding efficiencies. Then we extract the filter heralding efficiency by dividing out the heralding efficiency $\eta_{max,~s/i}$ when the filters are set to maximum bandwidth, which comes from nonunit coupling and detector efficiencies. Thus the filter heralding efficiencies are
\begin{align}
\eta_{f,s}= \frac{C}{S_i~\eta_{max,~s}},~~\eta_{f,i} = \frac{C}{S_s~\eta_{max,~i}}.
\end{align}
We confirmed that the peak filter transmission is independent of the WaveShaper's filter bandwidth assuring that the reduction in heralding efficiency is due to the fundamental tradeoff rather than technical imperfections (see plot in Supplemental Material).

\begin{figure}[t]
\centerline{\includegraphics[width=\linewidth]{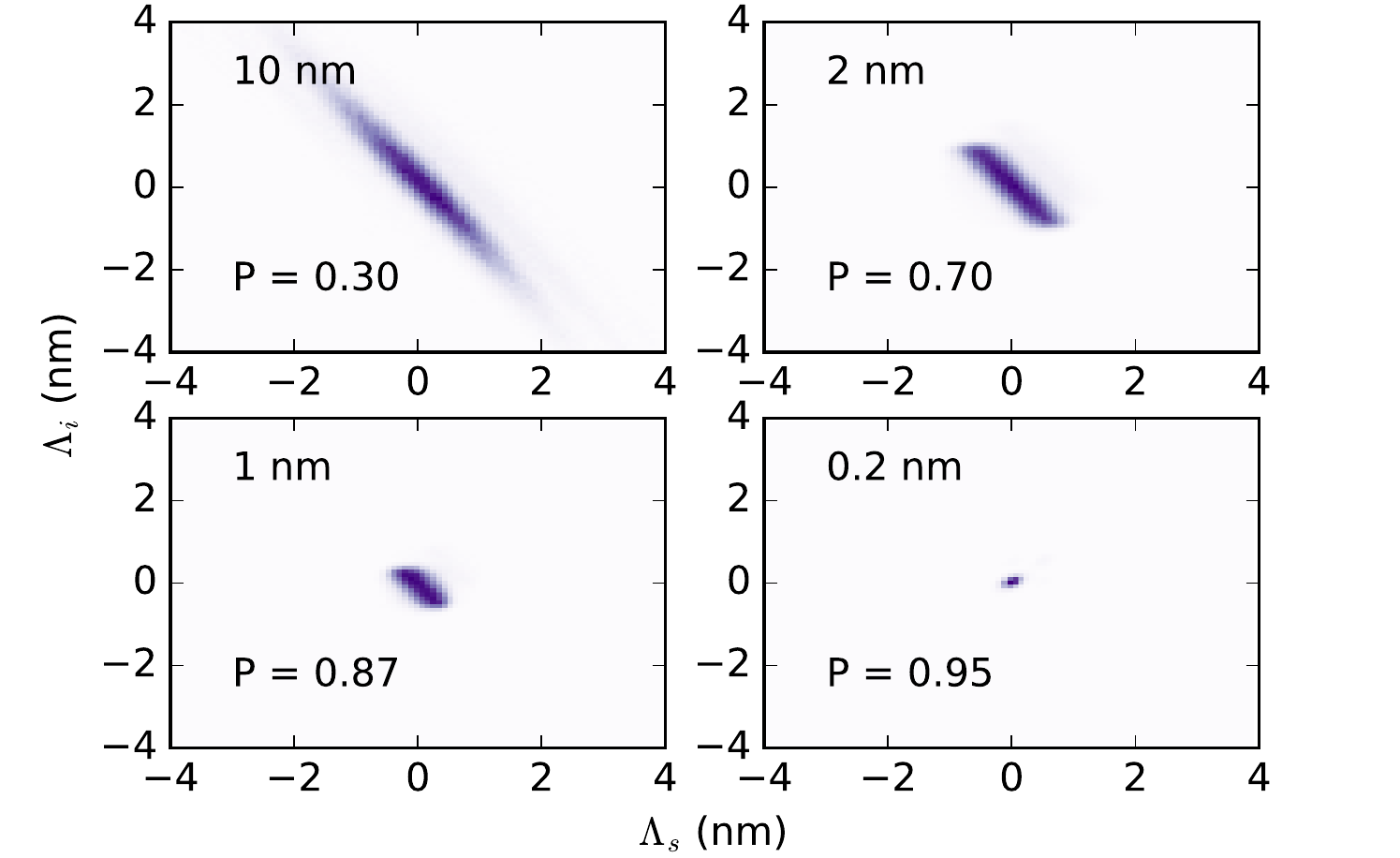}}
\caption{Measured joint spectral intensities of photon pairs, from nearly unfiltered (\SI{10}{\nano\meter} bandwidth) to strongly filtered and spectrally pure (\SI{0.2}{\nano\meter} bandwidth). The axis labels $\Lambda$ give the distance from the central wavelength.}
\label{fig.filter}
\end{figure}

We characterized the purity by measuring a joint spectral intensity with a time-of-flight spectrometer~\cite{Avenhaus:09}, assuming a constant phase of the joint spectrum~\cite{Gerrits:11}, and calculating $P = \mathrm{Tr}\left(\rho_a^2\right)$, where $\rho_a$ is the reduced spectral density matrix of the signal or idler photon~\cite{1367-2630-12-11-113052}. Using the JSI as an indicator of purity can be limited by artificial smoothing from limited spectrometer resolution and spectral phases that are not identifiable with intensity measurements. Thus we have employed as high a resolution as possible, and verified numerically that the expected phases due to pump chirp are negligible. 
We show in \cref{fig.filter} the joint spectral intensities after filtering and the corresponding purities, calculated with an additional time filter of twice the filter bandwidth to reduce technical noise from our laser's instability and limited timing resolution of our spectrometer.

\begin{figure}[h]
\centerline{\includegraphics[width=\linewidth]{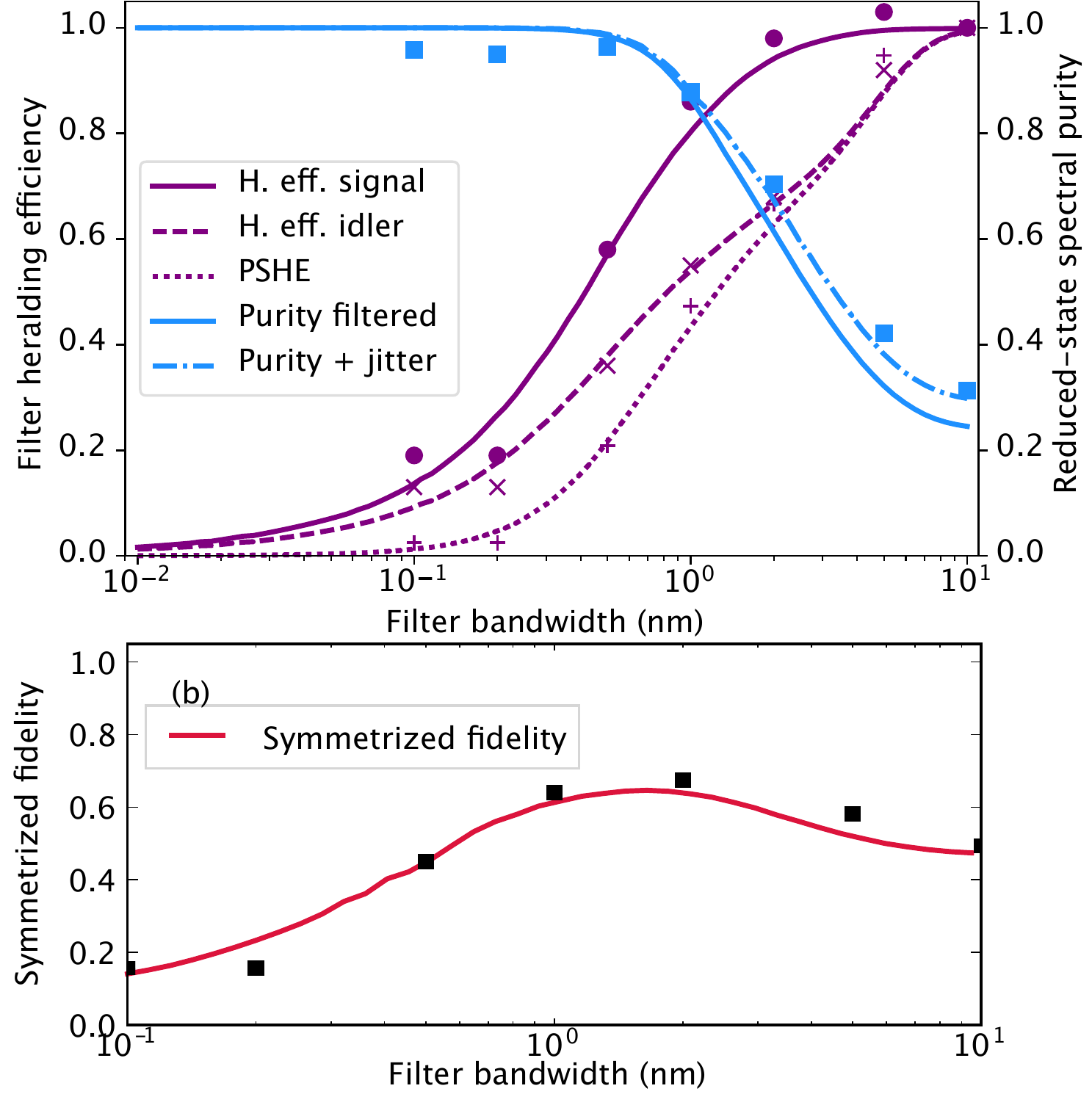}}
\caption{Experimental filter heralding efficiency, PSHE, and spectral purity (a) and symmetrized fidelity (b) versus filter bandwidth (points), with theoretical prediction (curves). The experimental data are shown with error bars from Poissonian statistics smaller than symbol size. Adding artificial jitter to the theoretical JSI makes the purity agree with the experiment for large filters, and thus this jitter has also been applied to the theoretical calculation for symmetrized fidelity.}
\label{fig.exp}
\end{figure}

The purity, filter heralding efficiencies, and symmetrized fidelity are plotted in \cref{fig.exp} and correspond reasonably well with the predictions after accounting for the asymmetry of our measured JSI. The limited resolution of our fiber spectrometer due to detector timing jitter tends to increase measured purities for large filters, as it rounds off sharp features of the JSI. Adding our experimental detector timing jitter of \SI{120}{\pico\second} to the theoretical JSI makes the predicted purity match the experiment for large filters. The remaining mismatch in the symmetrized fidelity could be due to small ripples in the WaveShaper transmission. The overall trend is clear: the increase in purity comes at a direct cost of heralding efficiency, and the fidelity of the signal and idler states to pure single photons cannot reach unity by filtering.

\section{Conclusion}
We have shown that spectral filtering of down-converted photons to increase the reduced-state spectral purity can lead to intrinsically low pair-symmetric heralding efficiencies, and cannot increase the symmetrized fidelity to a pure single photon beyond strong, general bounds. Our results suggest that, if high heralding efficiency of photon pairs is important, source engineering is required to generate spectrally decorrelated states, and for noise reduction only broadband filters should be used. The problem of reduced efficiency could also be avoided with carefully-designed cavities~\cite{Jeronimo-Moreno2010}, or more general time-frequency filtering~\cite{PhysRevA.82.043826} to directly select single spectral-temporal modes~\cite{Eckstein:11}. 

For example, without the reduction of heralding efficiency from narrowband filtering, the rate of 10-photon entanglement in two recent experiments~\cite{PhysRevLett.117.210502,Chen:17} could have been increased by a factor of 10 (counting only reduction of heralding efficiencies) or a factor 100 (counting all filtering losses).  For heralded photon sources, care must be taken when filtering the heralded photon so as not to decrease its heralding efficiency unnecessarily.  Finally, the analytic expressions we developed will be useful in designing the next generation of photon pair sources, as they allow optimization of the spectral purity and heralding efficiency with and without filtering. It would be interesting in future work to design the optimal filter shape that minimizes the purity-efficiency tradeoff, or maximizes the symmetrized fidelity.

\begin{acknowledgments}
We thank Vahid Ansari for providing WaveShaper code, and Viktor Quiring and Raimund Ricken for sample preparation. We acknowledge support from the Natural Sciences and Engineering Research Council of Canada, the Marie Curie Initial Training Network PICQUE (Photonic Integrated Compound Quantum Encoding, grant agreement no. 608062, funding Program: FP7-PEOPLE-2013-ITN, http://www.picque.eu), the DFG (Deutsche Forschungsgemeinschaft, grant no. SI 1115/4-1 and Gottfried Wilhelm Leibniz-Preis).
\end{acknowledgments}

\section*{Supplemental Material}
\subsection*{Analytic calculations of filter heralding efficiency, purity, and fidelity}
Here we find the filter heralding efficiency, purity, and fidelity of an arbitrary (but Gaussian) photon-pair joint spectrum, given Gaussian signal and idler filters of unit transmission and central frequencies that match the photons' central frequencies of $\omega_{s0/i0}$. The amplitude bandwidths for the pump ($\sigma_p$) and phasematching ($\sigma_{pm}$) are related to the FWHM intensity bandwidths that would be measured in the lab by 
\begin{align*}
&\sigma_p = \sigma_{p_{FWHM}}/(2\sqrt{2\ln2})\approx0.425\sigma_{p_{FWHM}},\\
&\sigma_{pm} = \sqrt{0.193}\sigma_{pm_{FWHM}}/(2\sqrt{2\ln2})\approx0.187\sigma_{pm_{FWHM}},
\end{align*}
where all bandwidths have units rad Hz. These can be converted to wavelength using the central pump wavelength for $\sigma_p$ and the central photon wavelength for $\sigma_{pm}$.

The filter heralding efficiency can be derived from the Klyshko heralding efficiency~\cite{0049-1748-10-9-A09} $\eta_{s/i} = \frac{C}{S_{i/s}}$, where $C$ are the number of coincidences and $S_{i/s}$ the number of singles in the idler/signal arm per integration time. Lumping all optical losses into $\eta_{opt}$ and keeping the filtering losses separate, the coincidences and singles are $C=n\eta_{opt}^2\Gamma_{both}$ and $S_{s/i}=n\eta_{opt}\Gamma_{s/i}$, where $n$ is the number of photon pairs produced, $\Gamma_{both}$ is the probability that both photons pass their respective filters, and $\Gamma_{s/i}$ is the probability that the signal/idler photon passes its filter. Thus the signal's filter heralding efficiency is 
\begin{align}
\eta_{f,s} = \frac{\eta_s}{\eta_{opt}}= \frac{\Gamma_{both}}{\Gamma_{i}},
\end{align}
and the idler's is 
\begin{align}
\eta_{f,i} = \frac{\eta_i}{\eta_{opt}}=\frac{\Gamma_{both}}{\Gamma_{s}}.
\end{align}

To calculate the heralding efficiency, we find the probability that both photons are passed by the filter, then the probability that each photon is passed individually.
The unfiltered state is given by $\ket{\psi} = \iint d\omega_s d\omega_i f\left(\omega_s, \omega_i\right)\ket{\omega_s}\ket{\omega_i}$, so the coincidence probability is 
\begin{align}
&\Gamma_{unfilt} = \left|\braket{\psi|\psi}\right|^2 \\ 
&= \iint d\omega_s d\omega_i \left|f\left(\omega_s, \omega_i\right) \right|^2\equiv 1.\nonumber
\end{align}
Following Ref.~\cite{0953-4075-46-5-055501}, we define $\Omega_s = \omega_s - \omega_{s0}$ and $\Omega_i = \omega_i - \omega_{i0}$ with $\omega_{s0}$ and $\omega_{i0}$ the respective central frequencies, approximate $\mathrm{sinc(x)}\approx\exp\left(-\alpha x^2\right)$ with $\alpha = 0.193$ in the joint spectral amplitude, add Gaussian filters with bandwidth $\sigma_{s/i} = \sigma_{s/i_{FWHM}}/(2\sqrt{2\ln2})$, and then neglect phase contributions, giving
\begin{align}
f\left(\omega_s, \omega_i\right)&\mathcal{F}_s(\omega_s)\mathcal{F}_i(\omega_i)\approx N \exp\left(-\frac{\left(\Omega_s+\Omega_i\right)^2}{4\sigma_p^2}\right)\\ 
&\times\exp\left(\frac{-\alpha\left(\Omega_s\sin{\theta} + \Omega_i\cos{\theta}\right)^2}{4\sigma_{pm}^2}\right)\\ \nonumber
&\times\exp\left(-\frac{\Omega_s^2}{4\sigma_s^2} - \frac{\Omega_i^2}{4\sigma_i^2}\right)\nonumber.
\end{align}
Collecting the terms in the exponentials gives
\begin{align}
f\left(\omega_s, \omega_i\right)&\approx N\exp\left(-\frac{a}{4}\Omega_s^2 - \frac{b}{4}\Omega_i^2 - \frac{c}{2}\Omega_s\Omega_i\right),
\end{align}
with~\cite{0953-4075-46-5-055501}
\begin{align*}
a&= \frac{\alpha^2\sin^2{\theta}}{\sigma_{pm}^2} + \frac{1}{\sigma_p^2} + \frac{1}{\sigma_s^2}\\ 
b&= \frac{\alpha^2\cos^2{\theta}}{\sigma_{pm}^2} + \frac{1}{\sigma_p^2} + \frac{1}{\sigma_i^2}\\ 
c&= \frac{\alpha^2\cos{\theta}\sin{\theta}}{\sigma_{pm}^2} + \frac{1}{\sigma_p^2}.
\end{align*}
Then without filters, with $\tfrac{1}{\sigma_s^2} = \tfrac{1}{\sigma_i^2}=0$ (and the corresponding $a_0$ and $b_0$), the coincidence probability is
\begin{align}
\Gamma_{unfilt} = N^2\iint d\Omega_s d\Omega_i\exp\left(-\frac{a_0}{2}\Omega_s^2 - \frac{b_0}{2}\Omega_i^2 - c\Omega_s\Omega_i\right).
\end{align}
This integral can be evaluated using the multi-dimensional generalization of a Gaussian function with
\begin{align}
A = \begin{pmatrix}
a_0 & c\\
c & b_0
\end{pmatrix}
\end{align}
 as
 
\begin{align}
\Gamma_{unfilt} = N^2 \frac{2\pi}{\sqrt{a_0b_0-c^2}},
\end{align}
giving the normalization 
\begin{align}
N^2 = \frac{\sqrt{a_0b_0-c^2}}{2\pi}.
\end{align}

Now, the coincidence count probability with signal and idler filters is
\begin{align}
\Gamma_{both} &= N^2\iint d\Omega_s d\Omega_i\exp\left(-\frac{a}{2}\Omega_s^2 - \frac{b}{2}\Omega_i^2 - c\Omega_s\Omega_i\right)\\ \nonumber
&= \sqrt{\frac{a_0b_0-c^2}{ab-c^2}}.
\end{align}

To find the marginal probabilities, we just set one of the filters to infinite bandwidth. Thus
\begin{align}
\Gamma_{i} =  \sqrt{\frac{a_0b_0-c^2}{a_0b-c^2}}\\
\Gamma_{s} =  \sqrt{\frac{a_0b_0-c^2}{ab_0-c^2}}.
\end{align}
Finally, the filter heralding efficiencies are
\begin{align}
\eta_{f,~s} = \frac{\Gamma_{both}}{\Gamma_{i}} = \sqrt{\frac{a_0b-c^2}{ab-c^2}},
\end{align}
and
\begin{align}
\eta_{f,~i} = \frac{\Gamma_{both}}{\Gamma_{s}} = \sqrt{\frac{ab_0-c^2}{ab-c^2}}.
\end{align}

To find the purity we need the reduced density matrix for signal or idler. Since our photons are entangled only in this spectral degree of freedom and we only consider the case when both photons are detected, they will have the same purity, and either reduced density matrix will suffice. We consider here only the spectral purity, neglecting vacuum and higher-order photon components. We find
\begin{align}
&\rho_s = \mathrm{Tr}_i\left(\ket{\psi}\bra{\psi}\right)\\
&= \iiint d\Omega_i d\Omega_s d\Omega_s^\prime f\left(\Omega_s, \Omega_i\right)f^*\left(\Omega_s^\prime, \Omega_i\right)\ket{\Omega_s}\bra{\Omega_s^\prime}. \nonumber
\end{align}
Then the purity is 
\begin{align}
P &= \mathrm{Tr}\left(\rho_s^2\right)\\\nonumber
 &= \iiiint d\Omega_s d\Omega_s^\prime d\Omega_i d\Omega_i^{\prime}\\\nonumber
 &\phantom{{}=1} \times f \left(\Omega_s, \Omega_i\right)f^* \left(\Omega_s^{\prime}, \Omega_i\right) f\left(\Omega_s^{\prime}, \Omega_i^\prime\right) f^*\left(\Omega_s, \Omega_i^{\prime}\right)\\ \nonumber
&= N^4\frac{(2\pi)^2}{\sqrt{a^2b^2-abc^2}}\\ \nonumber
&= \sqrt{\frac{\left(ab-c^2\right)^2}{a^2b^2-abc^2}}\\\nonumber
&= \sqrt{\frac{ab-c^2}{ab}}\nonumber
\end{align}

How do the purity and heralding efficiency depend on each other? To achieve high heralding efficiency requires $ab\rightarrow a_0b_0$, i.e. no filtering. To achieve high purity requires either $c=0$, i.e. source engineering to bring $\theta\in\left[\SI{90}{\degree},\SI{180}{\degree}\right]$ and matching the interaction length and pump bandwidth, or $ab\gg c^2$, i.e. strong filtering. But for $ab\gg c^2$, at least one of $a\gg |c|$ or $b\gg |c|$, implying at least one heralding efficiency tending to zero.

To find the symmetized fidelity we first consider the fidelity of the signal photon to an arbitrary Gaussian pure single photon state, after filtering and heralding by the (filtered) idler photon. The pure state is 
\begin{align}
\ket{1_p} = \int d\Omega g(\Omega)\ket{\Omega},
\end{align}
with
\begin{align}
 g(\Omega) = \left(\frac{d}{2\pi}\right)^{\frac{1}{4}} \exp\left(-\frac{d}{4}\Omega^2\right)
\end{align}
The fidelity (in the sense of probabilities~\cite{Jozsa:1994aa}) is 
\begin{align}
F_s=\bra{1_p}\rho\ket{1_p}
\end{align}
where $\rho = \left(1-\eta_{f,s}\right)\ket{0}\bra{0} + \eta_{f,s}\rho_s$, giving
\begin{align}
F_s &= \eta_{f,s}\iiint d\Omega_s d\Omega_s^\prime d\Omega_i \\\nonumber
&\times f \left(\Omega_s, \Omega_i\right)f^* \left(\Omega_s^{\prime}, \Omega_i\right)g(\Omega_s)g(\Omega_s^\prime)\\\nonumber
&= \eta_{f,s}\sqrt{\frac{4(ab-c^2)d}{b(a+d)^2-c^2(a+d)}}.
\end{align}
Differentiating with respect to $d$ to find the state which maximizes the fidelity gives
\begin{align}
d=\sqrt{\frac{a(ab-c^2)}{b}},
\end{align}
for the maximum fidelity
\begin{align}
F_s &= \frac{2\eta_{f,s}}{1+\sqrt{\frac{ab}{ab-c^2}}}\\\nonumber
&= \frac{2\eta_{f,s}P}{1+P}\\\nonumber
&= \frac{2\sqrt{a_0b-c^2}}{\sqrt{ab-c^2}+\sqrt{ab}}.
\end{align}
A similar procedure for the idler yields
\begin{align}
F_i= \frac{2\sqrt{ab_0-c^2}}{\sqrt{ab-c^2}+\sqrt{ab}}.
\end{align}
Combining these for the symmetrized efficiency gives
\begin{align}
F = \sqrt{F_sF_i} = \sqrt{\eta_{f,s}\eta_{f,i}}\frac{2P}{1+P}.
\end{align}

Finally we consider the purity-efficiency factor~\cite{PhysRevA.91.013819} of both photons together, which allows analytic optimization over the filter bandwidths. The factor is
\begin{align}
PEF&=\sqrt{P\eta_{f,s}\times P\eta_{f,i}} = \left(\frac{\left(a_0b-c^2\right)\left(ab_0-c^2\right)}{a^2b^2}\right)^{\frac{1}{4}}\\\nonumber
&=F\frac{1+P}{2}.
\end{align}
For $c\neq0$, the PEF can have in the best case any two of $\left\{P,\eta_{f,s},\eta_{f,i}\right\}$ approach 1, while the other approaches 0. For the phasematching angles $\theta\in\left[\SI{90}{\degree},\SI{180}{\degree}\right]$ one can have $c\rightarrow0$ allowing a PEF of 1. 
When $c^2>\tfrac{a_0b_0}{2}$, which corresponds to $\theta\in\left(\SI{15}{\degree},\SI{75}{\degree}\right)$, the maximum value of the PEF after optimizing the filter bandwidth can be found as
\begin{align}
PEF_{max}=\sqrt{\frac{a_0b_0}{4c^2}},
\end{align}
with optimal filter bandwidths defined by $a=\tfrac{2c^2}{b_0}$ and $b=\tfrac{2c^2}{a_0}$.
In the special cases of $\theta=\SI{45}{\degree}$ or for narrowband pumps or phasematching 
the PEF is upper-bounded by $\tfrac{1}{2}$ since $c^2\rightarrow a_0b_0$. The upper bound of the PEF for other phasematching angles depends on the angle and the pump and phasematching bandwidths, but is in general $<\tfrac{1}{\sqrt{2}}$ for $\theta\in\left(\SI{15}{\degree},\SI{75}{\degree}\right)$.

\subsection*{WaveShaper Transmision}
\begin{figure}[t]
\centerline{\includegraphics[width=\linewidth]{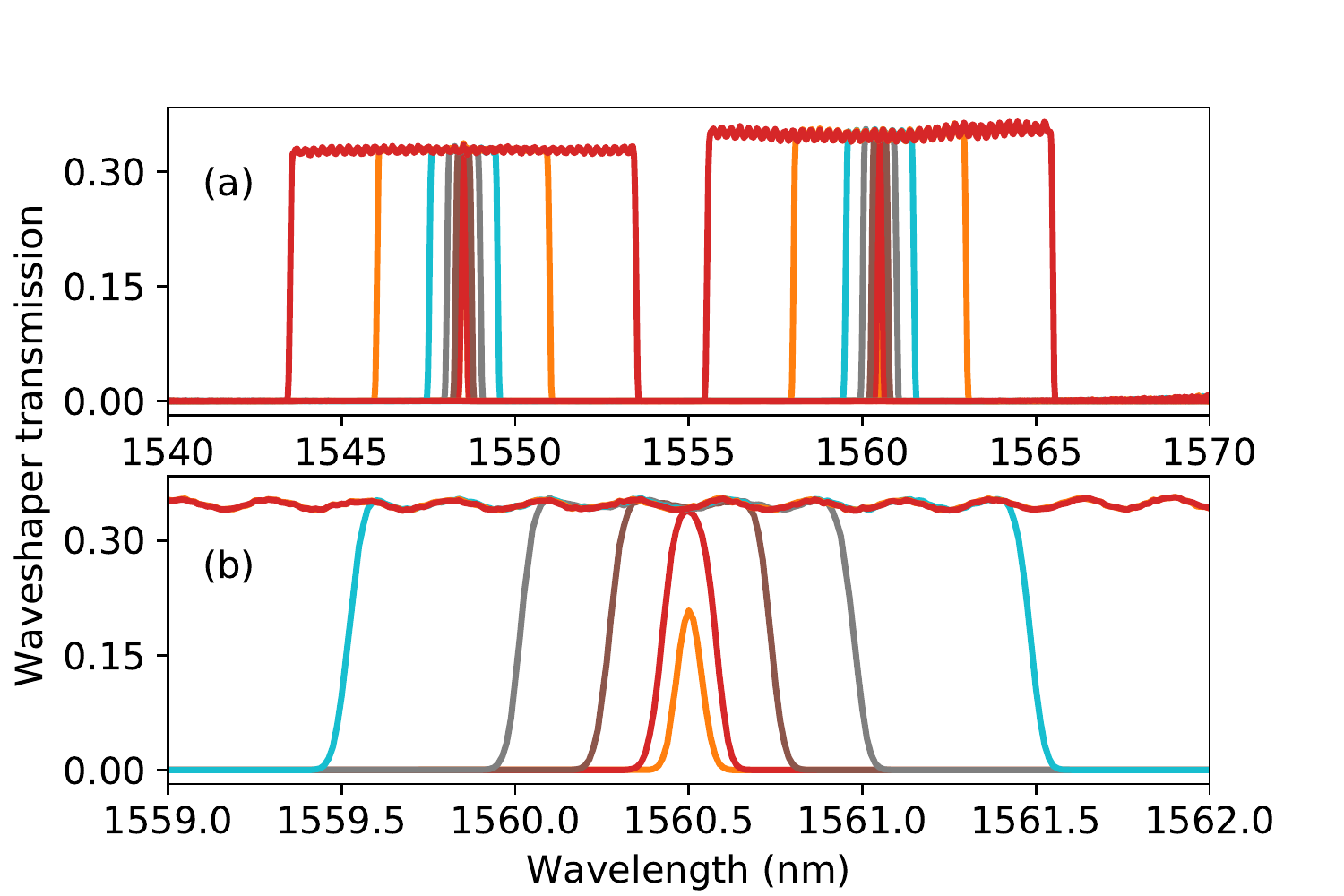}}
\caption{Measured transmission of the tunable filters for various square filter settings. The full bandwidth with the signal and idler filters is shown in corresponding colors in (a), while (b) shows a zoom of the signal filters, showing no change in peak transmission until $<\SI{0.2}{\nano\meter}$, which is limited by the resolution of our spectrometer.}
\label{fig.waveshaper}
\end{figure}

In the experiment, we confirmed that the spectral filters applied are nearly square with direct measurements of the WaveShaper, shown in \cref{fig.waveshaper}. The transmission loss is about \SI{4.8}{\decibel} for the signal photon and \SI{4.4}{\decibel} for the idler.

\bibliography{spdc_filtering}

\end{document}